\documentclass[english,twocolumn]{revtex4}
\usepackage[T1]{fontenc}
\usepackage[latin9]{inputenc}
\usepackage{verbatim}
\usepackage{amsbsy}
\usepackage{amstext}
\usepackage{amssymb}
\usepackage{graphicx}
\usepackage{esint}

\makeatletter

\@ifundefined{textcolor}{}
{%
 \definecolor{BLACK}{gray}{0}
 \definecolor{WHITE}{gray}{1}
 \definecolor{RED}{rgb}{1,0,0}
 \definecolor{GREEN}{rgb}{0,1,0}
 \definecolor{BLUE}{rgb}{0,0,1}
 \definecolor{CYAN}{cmyk}{1,0,0,0}
 \definecolor{MAGENTA}{cmyk}{0,1,0,0}
 \definecolor{YELLOW}{cmyk}{0,0,1,0}
}

\usepackage[percent]{overpic}
\makeatother

\usepackage{babel}
\begin{document}
\global\long\def\braketop#1#2#3{\left\langle #1\vphantom{#2}\vphantom{#3}\left|#2\vphantom{#1}\vphantom{#3}\right|#3\vphantom{#1}\vphantom{#2}\right\rangle }
\global\long\def\braket#1#2{\left\langle \left.#1\vphantom{#2}\right|#2\right\rangle }
\global\long\def\bra#1{\left\langle #1\right|}
\global\long\def\ket#1{\left|#1\right\rangle }
\global\long\def\ketbra#1#2{\left|#1\vphantom{#2}\right\rangle \left\langle \vphantom{#1}#2\right|}
\global\long\def\valley{K'}
\global\long\def\othervalley{K}
\global\long\def\figlab#1{(#1)}
\global\long\def\labref#1#2{#1.\sim#2}
\global\long\def\vec#1{\mathbf{#1}}

\title{Husimi Maps in Lattices}

\date{06/05/12}

\author{Douglas J. Mason, Mario F. Borunda, and Eric J. Heller}

\affiliation{Department of Physics, Harvard University, Cambridge, MA 02138, USA}
\begin{abstract}
We build upon previous work that used coherent states as a measurement
of the local phase space and extended the flux operator by adapting
the Husimi projection to produce a vector field called the Husimi
map. In this article, we extend its definition from continuous systems
to lattices. This requires making several adjustments to incorporate
effects such as group velocity and multiple bands. Several phenomena
which uniquely occur in lattice systems, like group-velocity warping
and internal Bragg diffraction, are explained and demonstrated using
Husimi maps. We also show that scattering points between bands and
valleys can be identified in the divergence of the Husimi map.
\end{abstract}
\maketitle

\section{Introduction}

In Mason \emph{et al.}\cite{Mason-PRL}, we introduced a new interpretation
of the probability flux operator 
\begin{equation}
\hat{\vec j}(\vec r)=\frac{1}{2m}\left(\ketbra{\vec r}{\vec r}\hat{p}+\hat{p}\ketbra{\vec r}{\vec r}\right)\label{eq:Flux-Operator}
\end{equation}
by expressing its eigenstates as the limit of measurements by infinitesimally
small coherent states. Our approach yields a new perspective on flux
measurements and provides a novel tool for visualizing wavefunctions
which parallel the probability flux map. Because they are based on
the Husimi projection technique\cite{Husimi}, these visualizations
are called \textquotedbl{}Husimi maps\textquotedbl{}. Husimi maps
improve the understanding of the semiclassical paths underlying the
quantum wavefunctions and can be of use even for systems where the
traditional flux has little success (i.e., when it is either zero
or strongly misleading). Later work\cite{Mason-Husimi-Continuous}
further developed the numerical framework of the Husimi map and applied
it to a wider variety of systems by incorporating local potentials
and examining flux through open systems.

This article expands the Husimi map technique from continuous, free-particle
systems like the two-dimensional electron gas (2DEG) to crystalline
systems like graphene. While the extended wavefunction of an electron
in a crystal is continuous, the potential imposed by the nuclei can
be modeled by replacing the continuum with localized wavefunctions
centered at individual tight-binding lattice sites. These individual
wavefunctions combine to form a model of the entire wavefunction,
which now defines their envelope function. In this model, Eq.~\ref{eq:Flux-Operator}
describes not the probability flow at an infinitesimal point, but
the flow of probability in and out of the localized wavefunction at
a single site.

Lattice systems can behave very differently from continuous systems.
For instance, the orientation of the group velocity vector, which
dictates classical dynamics, can strongly diverge from the wavevector,
which was the initial foundation of the Husimi projection. In fact,
the group-velocity space can be so strongly restricted that classical
trajectories are only permitted along certain directions, dramatically
affecting the dynamics of states that inhabit lattice systems. When
these trajectories hit a boundary, internal Bragg diffraction can
produce additional nonclassical ray reflections.

Here we explore two-dimensional square and honeycomb lattices; extension
to three-dimensional systems is straightforward. Honeycombs induce
an additional phenomenon: the presence of multiple bands and valleys,
by which different quasiparticles can propagate and interfere. While
the flux operator is unable to reflect any of these behaviors, in
this article we show that with proper modifications, the Husimi projection
can handle them with ease.

This paper is organized as follows: In Section \ref{sub:Husimi-Projection}
we provide the definition of the Husimi projection for continuous
system and then modify the Husimi projection in Section \ref{sub:Group-Velocity-Method}
to represent the group velocity and multiple bands. In Section \ref{sub:Billiard-Eigenstates},
we apply the Husimi projection to square lattices near the band center
where group-velocity effects are strongest, and in Section \ref{sub:Graphene-Eigenstates},
we examine the graphene honeycomb lattice. Finally, in Section \ref{sub:Internal-Bragg-Diffraction},
we provide an interpretation of unusual boundary reflections found
in Sections \ref{sub:Billiard-Eigenstates} and \ref{sub:Graphene-Eigenstates}
by demonstrating and measuring internal Bragg diffraction.

\section{Method}

\subsection{Definition of the Husimi Projection\label{sub:Husimi-Projection}}

Building off work in Husimi\cite{Husimi} and Mason \emph{et al.}\cite{Mason-Husimi-Continuous},
we define the Husimi function as a measurement between a wavefunction
$\psi\left(\left\{ \vec r_{i}\right\} \right)$ and a coherent state
$\ket{\vec r_{0},\vec k_{0},\sigma}$, which minimizes joint uncertainty
in spatial and momentum coordinates. For lattice systems, the wavefunction
represents the probability amplitude multiplier of localized wavefunctions
indexed at discrete lattice sites, which are associated with discrete
positions in the set $\left\{ \vec r_{i}\right\} $. The coherent
state is also an envelope function over localized wavefunctions, defined
by the Gaussian function
\[
e^{-\left(\vec r_{i}-\vec r_{0}\right)^{2}/4\sigma^{2}+i\vec k_{0}\cdot\vec r_{i}}
\]
 centered around $\vec r_{0}$ and $\vec k_{0}$. The parameter $\sigma$
defines the spatial spread of the coherent state and the uncertainties
in space and momentum according to the well-known relation
\begin{equation}
\Delta x\propto\frac{1}{\Delta k}\propto\sigma.
\end{equation}
As a result, there is a trade-off for any value of $\sigma$ selected:
for small $\sigma$, spatial resolution is improved at the expense
of resolution in momentum space, and vice versa for large $\sigma$. 

We can explicitly write out the projection of the wavefunction onto
the coherent state as 

\begin{eqnarray}
\braket{\psi}{\vec r_{0},\vec k_{0},\sigma} & = & \left(\frac{1}{\sigma\sqrt{\pi/2}}\right)^{d/2}\nonumber \\
 &  & \times\sum_{i}\psi\left(\vec r_{i}\right)e^{-\left(\vec r_{i}-\vec r_{0}\right)^{2}/4\sigma^{2}+i\vec k_{0}\cdot\vec r_{i}},\label{eq:Gaussian-Function}
\end{eqnarray}
where $d$ is the number of dimensions in the system. The Husimi function
is then defined as
\begin{equation}
\text{Hu}\left(\vec r_{0},\vec k_{0},\sigma;\psi\left(\left\{ \vec r_{i}\right\} \right)\right)=\left|\braket{\psi}{\vec r_{0},\vec k_{0},\sigma}\right|^{2}.\label{eq:Husimi}
\end{equation}
If we weight the Husimi function by the central wavevector $\vec k_{0}$,
we obtain the Husimi vector. When momentum space is explored at a
point by many Husimi vectors, the result is the full Husimi projection.

If all the Husimi vectors at a point are summed, the Husimi function
can be used to construct and generalize the flux operator, resulting
in the vector-valued function $\vec{Hu}\left(\vec r_{0},\sigma;\psi\left(\left\{ \vec r_{i}\right\} \right)\right)$
equal to 
\begin{equation}
\vec{Hu}\left(\vec r_{0},\sigma;\psi\left(\left\{ \vec r_{i}\right\} \right)\right)=\int\left|\braket{\psi}{\vec r_{0},\vec k_{0},\sigma}\right|^{2}\vec k_{0}dk_{0}.\label{eq:Husimi-Vector-Old}
\end{equation}
Earlier work has shown that for $\sigma k\ll1$, Eq.~\ref{eq:Husimi-Vector-Old}
is proportional to the traditional flux vector expectation value \cite{Mason-Husimi-Continuous}.
For lattice systems, the traditional flux becomes a finite-difference
approximation defined by the lattice.%

\subsection{The Hamiltonian\label{sub:The-Hamiltonian}}

This paper examines Hamiltonians using the nearest-neighbor tight-binding
approximation 
\begin{equation}
H=\sum_{i}\epsilon_{i}\mathbf{a}_{i}^{\dagger}\mathbf{a}_{i}-t\sum_{\left\langle ij\right\rangle }\mathbf{a}_{i}^{\dagger}\mathbf{a}_{j}
\end{equation}
where $\mathbf{a}_{i}^{\dagger}$ is the creation operator at orbital
site $i$ and we sum over the set $\left\langle ij\right\rangle $
of nearest neighbors. The quantity $\epsilon_{i}$ is the on-site
energy and $t$ is the hopping energy scale. For the square lattices,
we set $\epsilon=-4t$. For systems at energies $E<0.5t$, the tight-binding
Hamiltonian is a close approximation to the effective mass envelope
function Hamiltonian $H=-\frac{p^{2}}{2m}+U(\vec r)$ where $t=\hbar^{2}/(2m^{\ast}a^{2})$
and $a$ is the mesh lattice spacing. For the honeycomb lattice, parameters
are set to the common values in the literature for graphene: $\epsilon=0$
and $t=2.7\text{eV}$\cite{Geim-Nature,castro-neto}. Eigenstates
of closed stadium billiard systems are computed using the standard
sparse matrix eigensolvers. %

\subsection{Group Velocity\label{sub:Group-Velocity-Method}}

In the original introduction of the Husimi map\cite{Mason-Husimi-Continuous},
each Husimi function is weighted by the wavevector of the coherent
state to produce a visual guide to the classical dynamics of the system.
Summing all the vectors equates to the flux operator (Eq.~\ref{eq:Husimi-Vector-Old})
when the coherent states are sufficiently small. This equivalence
holds in lattices, however, the direction and magnitude of the group
velocity $\vec{\nabla}_{\vec k}E\left(\vec k\right)$ can strongly
diverge from the wavevector. Since a coherent state, which is now
defined as an envelope function over localized wavefunctions, follows
the group-velocity vector instead of its wavevector, it is necessary
to weight the Husimi function by group-velocity vectors to indicate
the classical dynamics. As a result, the Husimi projection indicates
the classical flow of quasiparticles, in contrast to the flux operator
(Eq.~\ref{eq:Flux-Operator-1}), which instead indicates the flow
of probability.

\begin{figure}

\begin{centering}
\includegraphics[width=0.8\columnwidth]{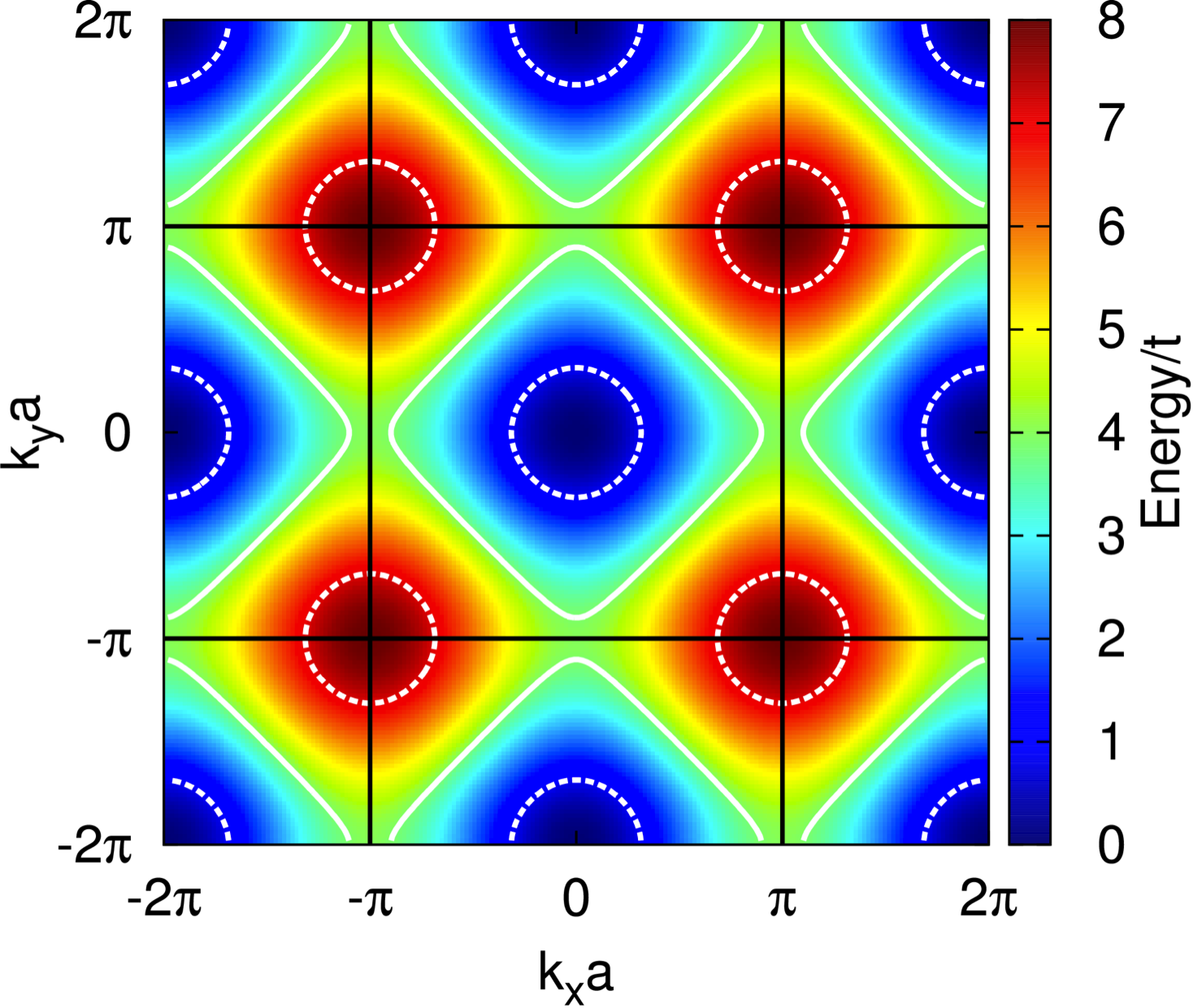}
\par\end{centering}

\begin{centering}
\caption{\label{fig:Dispersion Relation}The two-dimensional dispersion relation
for the square lattice demonstrates strong group-velocity warping
at the band center ($E=4t$). The dispersion relations for $E=$ $0.9t$,
and $7.1t$ (dashed white lines) are nearly circular, while the relation
near the band edge at $E=3.9t$ (solid white lines) shows strong warping.}

\par\end{centering}

\end{figure}
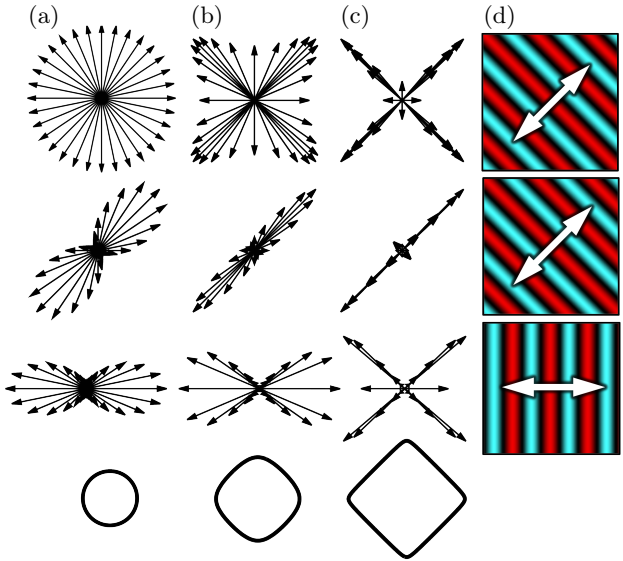
\begin{figure}
\begin{centering}
\begin{overpic}[width=1\columnwidth]{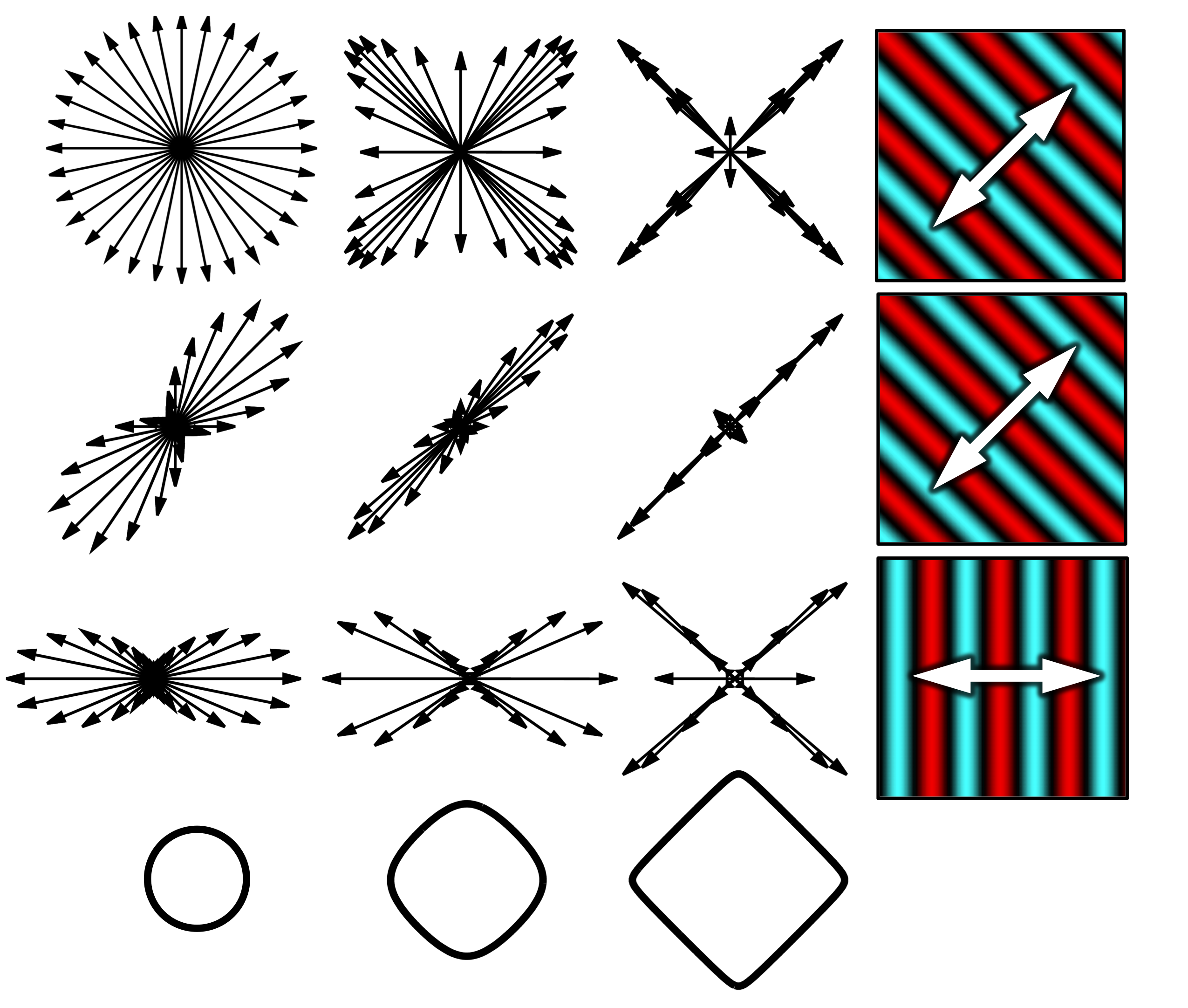}\put(4,84){\figlab{a}}\put(29,84){\figlab{b}}\put(52,84){\figlab{c}}\put(74,84){\figlab{d}}\end{overpic}
\par\end{centering}

\caption{\label{fig:Sunbursts}The group-velocity Husimi projection for the
square lattice is strongly affected by warping at energies near the
band center (Fig.~\ref{fig:Dispersion Relation}). Husimi projections
are shown for the square lattice for the group-velocity representation
at $E=0.9t$(a), $3.0t$(b), and $3.9t$(c) with relative uncertainties
of $\Delta k/k=2$ (top) and $50\%$ (middle and bottom). A schematic
of the dispersion relation contour at each energy is shown at the
far bottom. The generating wavefunction $\psi$ for each row is shown
in (d). In the top and middle row the test wavefunction is a cosine
wave pointing along the $45^{\circ}$ diagonal, and in the bottom
along the $0^{\circ}$ horizontal. }
\end{figure}

At low energies, the square lattice closely approximates a free-particle
continuous system so that this modification is minimal. At higher
energies, however, the mapping from the wavevector to group velocity
can be strongly constricted. For example, at energies near the band
center of $E=4t$, there are only four directions available to the
group velocity in the square lattice, as shown in the solid white
contour in Fig.~\ref{fig:Dispersion Relation} at $E=3.9t$.

To visualize this effect, we show group-velocity Husimi projections
at three representative energies in Fig.~\ref{fig:Sunbursts} for
the square lattice. Thirty-two equally-spaced angles along a circle
are chosen to represent the local momentum space. Wavevectors are
chosen with these angles to satisfy the dispersion relation for a
given energy. 

At energies away from the band center for the square lattice, semi-classical
trajectories can assume any direction, but near the band center they
must follow preferred directions determined by the group-velocity
warping. However, the manner in which they do so may differ. This
can be seen in Fig.~\ref{fig:Sunbursts} which examines two cosine-wave
states with different wavevectors. As the energy of the system increases
from left-to-right, group-velocity warping draws Husimi vectors, and
the classical paths, towards four preferred directions. When the generating
wavevector points along one of these directions, group-velocity warping
merely sharpens the profile. When the generating wavevector points
\emph{in between }the preferred directions, as in the bottom row of
Fig.~\ref{fig:Sunbursts}, the classical trajectories are more strongly
dependent upon the system energy.

Any expectation value over a wavefunction must be evaluated, usually
by an integral, over a complete basis. As a result, any expectation
value derived from the Husimi projection must be first computed from
the wavevector basis; modifications to account for group velocity
are determined afterwards. For instance, in the Multi-Modal Algorithm,
which approximates the full Husimi projection by a subset of local
plane waves (see Mason \emph{et al.}\cite{Mason-Husimi-Continuous}
for more details), each approximation is achieved by computing the
dot product between the Husimi projection and template projections
in $k$-space. Because of group-velocity warping, the resulting wavevectors
no longer indicate classical flow. To address this problem, the resulting
dominant wavevectors are then mapped onto group velocity by taking
the local derivative of the dispersion relation.

\subsection{Band Structure}

The number of bands for a lattice system is equal to the number of
tight-binding orbitals in the unit cell\cite{ashcroft-and-mermin}.
The square lattice has only one unique tight-binding orbital and only
one band, but due to the warping in the band structure, distinct behaviors
result at energies above $E=4t$, corresponding to the hole pocket
(see the contour lines in Fig.~\ref{fig:Dispersion Relation} near
the corners of the Brillouin zone). However, because the quadratic
dispersion relations at $E=0$ and $E=8t$ are separated by energy,
a semi-classical interpretation of a wavefunction is always constrained
to one relation or the other.

\begin{figure}

\begin{centering}
\includegraphics[width=0.85\columnwidth]{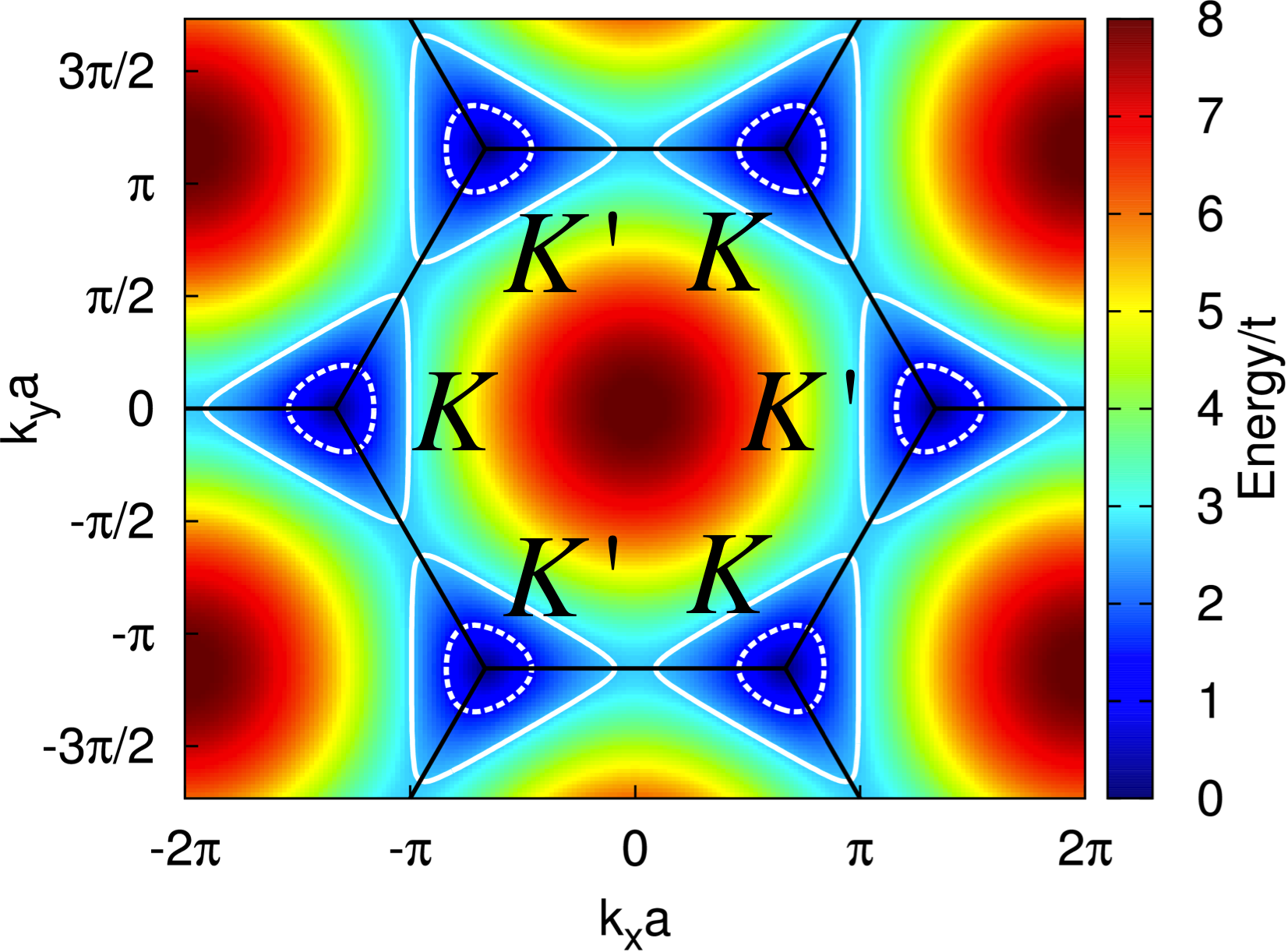}
\par\end{centering}

\caption{\label{fig:Dispersion Relation-Graphene}Like the square lattice in
Fig.~\ref{fig:Dispersion Relation}, the two-dimensional dispersion
relation for the honeycomb lattice demonstrates strong group-velocity
warping at energies away from the Dirac point. The dispersion relation
for $E=0.5t$ (dashed white lines) is nearly circular, while the relation
at $E=0.98t$ (solid white lines) shows strong warping. The $K$ and
$K'$ valleys are indicated.}
\end{figure}

In the honeycomb lattice, however, there are \emph{two} unique orbitals
in the lattice structure, yielding two bands that touch at the Dirac
point at $E=0t$. But more interestingly, the band structure warps
each band to produce the inequivalent $K$ and $K^{\prime}$ valleys
at the Dirac point, which are indicated in Fig.~\ref{fig:Dispersion Relation-Graphene}.
Unlike the square lattice, these two valleys co-exist in the energy
range $-t<E<t$. 

These valleys exhibit a linear dispersion relation near the Dirac
point\cite{Geim-Nature}. At energies away from the Dirac point, the
two valleys undergo group-velocity warping that emphasizes three directions,
which is referred to as ``trigonal warping''\cite{castro-neto}.
The effects of trigonal warping can be significant even at energies
as low as $0.2t$. 

The Husimi map can assist visualizations of intervalley scattering,
the scattering of quasiparticles between the two valleys which are
part of the same band. To resolve the two valleys, it is simply necessary
that the uncertainty of the coherent state is small enough in $k$-space
to unambiguously resolve the wavevectors of each valley. Because the
two valleys of graphene are well-separated and only come close at
the corners of each triangle in Fig.~\ref{fig:Dispersion Relation-Graphene},
the Husimi projection can clearly resolve the two valleys at most
energies. More complicated lattices can have additional bands, and
any automated method for calculating Husimi maps for these systems
have to take their mutual distance in $k$-space into account. 

\section{Results}

\subsection{Stadium Billiard Eigenstates of the Square Lattice\label{sub:Billiard-Eigenstates}}

\begin{figure}
\begin{centering}
\begin{overpic}[width=0.85\columnwidth]{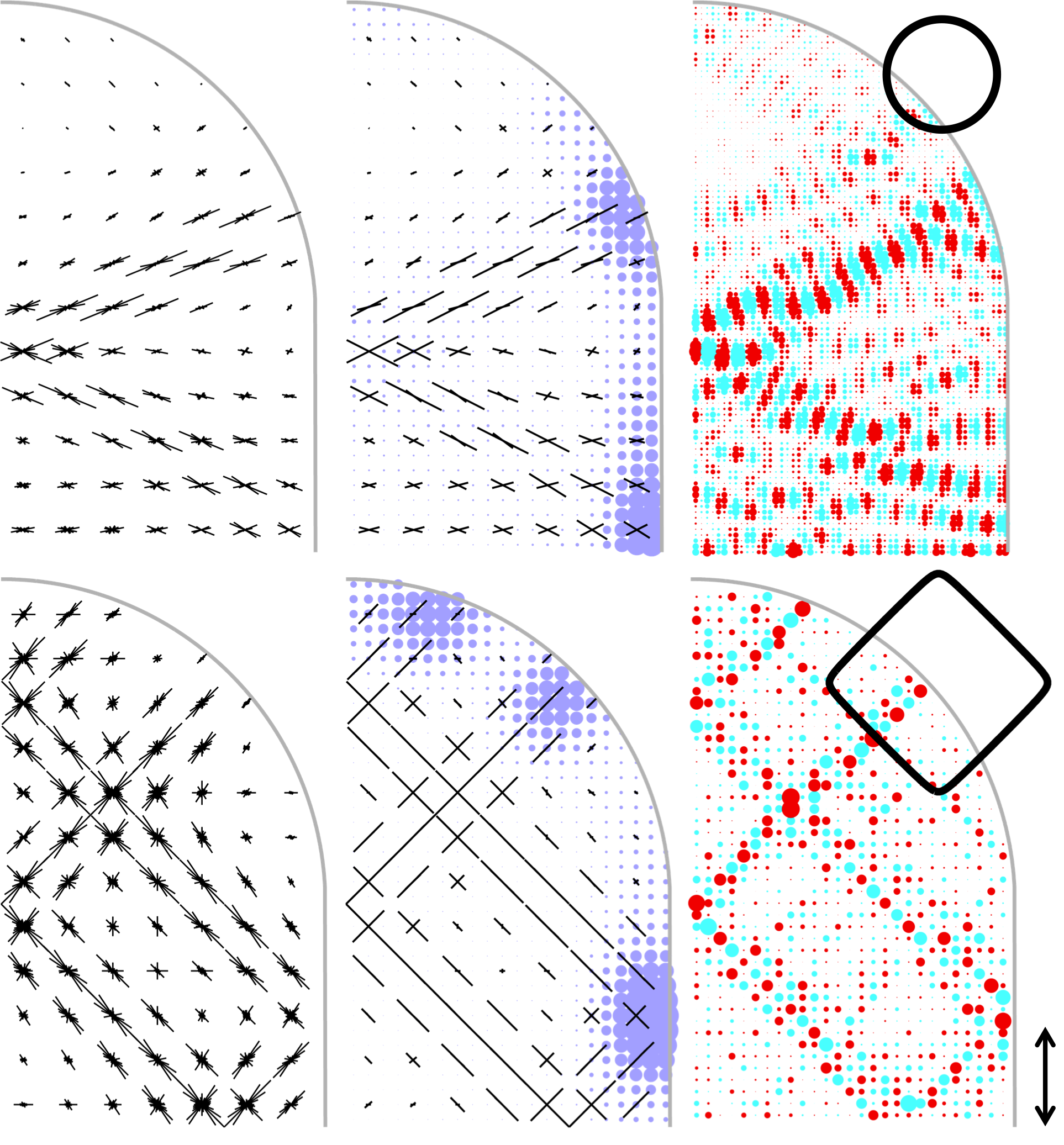}\put(-5,97){\figlab{a}}\put(-5,46){\figlab{b}}\end{overpic}
\par\end{centering}

\caption{\label{fig:Billiard Eigenstates} The full wavevector Husimi map (left),
multi-modal analysis (middle) and wavefunction (right) for two stadium
eigenstates at energies $E_{1}=1.496t$ (a) and $E_{2}=3.982t$ (b)
(a schematic of the dispersion relation contour at each energy is
shown in the insets). The uncertainty for each projection is set to
$\Delta k/k=10\%$, and the spread of the coherent state is indicated
by double arrows on the right. Angular deflection (Eq.~\ref{eq:Summed-Divergence})
is indicated in blue. Each eigenstate has similar characteristic wavelengths,
but the lower eigenstate is sampled with half the linear resolution,
causing its energy to go up and the group velocity to become more
restrictive. }
\end{figure}

In Fig.~\ref{fig:Billiard Eigenstates}, we examine two closed stadium
billiard systems with identical geometric parameters. Both systems
are created using the square-lattice tight-binding model, but the
lattice constant in Fig.~\ref{fig:Billiard Eigenstates}b is twice
as large, so the system possesses far fewer sampling points and experiences
stronger effects from group-velocity warping. These systems connect
to the lattice-sampled Schrodinger equation for a continuous system,
which can avoid the effects of group-velocity warping by increasing
the number of sample points in the system (See Fig.~\ref{fig:Billiard Eigenstates}).
However, lattice spacing is not an adjustable parameter in atomic
systems, and group velocity must be given careful attention.

Keeping the characteristic wavelength constant raises the energy in
any system with a longer lattice constant. In Fig.~\ref{fig:Billiard Eigenstates}b,
an eigenstate of the system is shown with energy $E_{2}=3.892t$,
near the band center. The energy for the system in Fig.~\ref{fig:Billiard Eigenstates}a
is chosen to reflect the same characteristic wavelength, which depends
upon which direction in $k$-space is considered. Along the $k_{x}$-axis,
the energy is bounded below by $\frac{E_{2}}{t_{2}}=-2\left(\cos\left(\frac{a_{2}}{a_{1}}\cos^{-1}\left[1-\frac{E_{1}}{2t_{1}}\right]\right)-1\right)$,
and at 45-degrees from the $k_{x}$-axis, it is bounded from above
by $\frac{E_{2}}{t_{2}}=-4\left(\cos\left(\frac{a_{2}}{a_{1}}\cos^{-1}\left[1-\frac{E_{1}}{4t_{1}}\right]\right)-1\right)$.
By setting $\frac{a_{2}}{a_{1}}=\frac{1}{2}$ and $t_{1}=t_{2}=t$
an eigenstate is chosen with an energy near the average of the bounds
at $E_{1}=1.496t$. 

The classical trajectories indicated by the Husimi map in the low-energy
state in Fig.~\ref{fig:Billiard Eigenstates}a point along directions
oblique to the $45^{\circ}$ diagonals and intermingle among other
paths at other angles. The Husimi map for the higher energy eigenstate
in Fig.~\ref{fig:Billiard Eigenstates}b instead \emph{only} indicates
classical trajectories along the $45^{\circ}$ diagonals. Moreover,
the trajectories in the higher-energy system are much clearer since
they are reinforced by a restricted group-velocity space.

The Husimi map makes it possible to measure ``angular deflection'',
which reflects how classical trajectories deviate from the straight
line in response to the system. Angular deflection thus provides a
map of where the boundaries or external potentials most strongly effect
these dynamics, and can be interpreted as a force on the particle
represented by the wavefunction.

For lattice systems, the original definition of angular deflection
provided in Mason \emph{et al.}\cite{Mason-Husimi-Continuous} must
be modified to account for group velocity. It can thus be defined
\begin{equation}
Q_{\text{ang.}}\left(\vec r;\Psi\right)=\int D_{\text{abs.}}(\vec r,\vec k;\Psi)\left|\boldsymbol{\nabla}_{\vec k}E\left(\vec k\right)\right|d^{d}k,
\end{equation}
where the quantity $\boldsymbol{\nabla}_{\vec k}E\left(\vec k\right)$
represents the group-velocity vector corresponding to the wavevector
$\vec k'$, and the integral covers all wavevectors satisfying the
dispersion relation. The quantity $D_{\text{abs.}}(\vec r,\vec k;\Psi)$
is defined as the Gaussian-weighted absolute divergence of the Husimi
function for one particular trajectory angle 
\begin{eqnarray}
D_{\text{abs.}}(\vec r,\vec k;\Psi) & = & \int\sum_{i=1}^{d}\left|\frac{\text{Hu}\left(\vec k,\vec r';\Psi\right)-\text{Hu}\left(\vec k,\vec r;\Psi\right)}{\left(\vec r'-\vec r\right)\cdot\vec e_{i}}\right|\nonumber \\
 &  & \times\exp\left[\frac{\left(\vec r'-\vec r\right)^{2}}{2\sigma^{2}}\right]d^{d}r',\label{eq:Summed-Divergence}
\end{eqnarray}
where we sum over the $d$ orthogonal directions each associated with
unit vector $\vec e_{i}$. 

Fig.~\ref{fig:Billiard Eigenstates} shows angular deflection in
blue, concentrated on the boundary as expected. Because the resolution
of angular deflection is limited by the spread of the coherent state
used for Husimi sampling, its spread into the bulk from the boundary
exhibits the same Gaussian distribution that is used for the test
wavepacket. It is worth noting that without the proper modifications,
angular deflection based on the wavevector shows non-trivial results
in the bulk of the system even when there are no external fields. 

This also suggests that modifications may be in order for other metrics
for lattice systems. For instance, by coordinating the boundary divergence
with each wavevector, one can compute the quantum analog of a state's
Poincare map\cite{gutzwiller-chaos}. In Birkhoff coordinates\cite{Birhoff,Birkhoff-User},
the angle of impact is mapped against a coordinate along the boundary,
and both fully quantum\cite{heller-more-husimi,Heller-billiard-with-Husimi}
and classical\cite{poincare-andreev} variations have become valuable
tools in quantum chaos. By incorporating group-velocity considerations,
these metrics may be extended to lattice systems.

\subsection{Stadium Billiard Eigenstates of the Honeycomb Lattice\label{sub:Graphene-Eigenstates}}

\begin{figure}
\begin{centering}
\begin{overpic}[width=0.85\columnwidth]{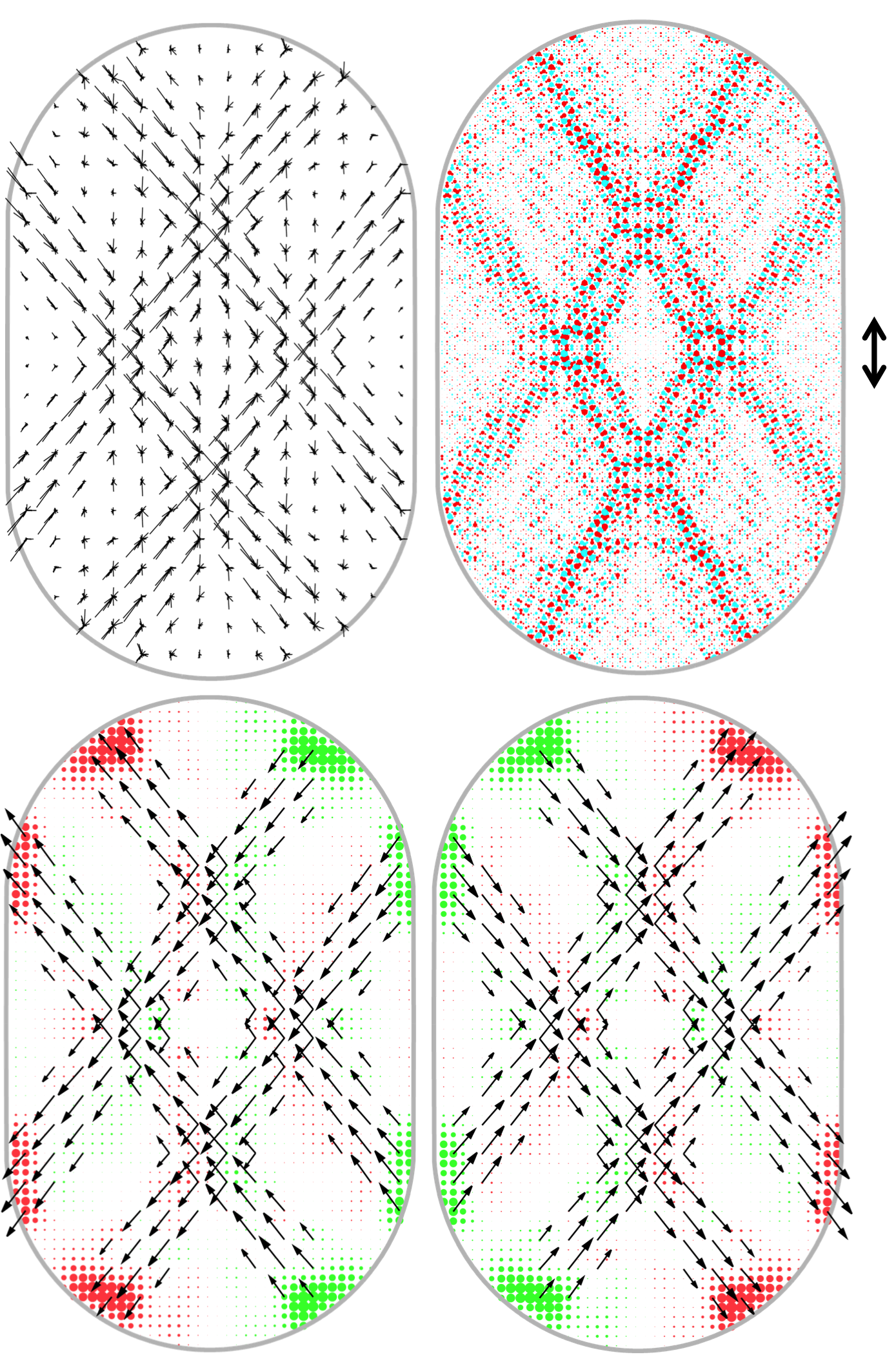}\put(0,96){\figlab{a}}\put(31,96){\figlab{b}}\put(0,47){\figlab{c}}\put(31,47){\figlab{d}}\end{overpic}
\par\end{centering}

\caption{\label{fig:Graphene-Eig}The full Husimi map around the $\valley$
valley(a), the wavefunction(b), the multi-modal analysis for the $\valley$
valley(c) and for the $\othervalley$ valley(d) for a high-energy
eigenstate of the honeycomb lattice at $E=0.786t$. This system is
a closed stadium billiard system with 20270 lattice points. The relative
uncertainty in all calculations is $\Delta k/k=20\%$ with the coherent
state spread indicated by the double-arrows. Because of time-reversal
symmetry, the Husimi maps in (c) and (d) are exact inverses of each
other. Unlike the square lattice, the summing the Husimi vectors for
each valley in a honeycomb lattice gives non-zero results for a closed
system, giving rise to non-trivial divergences along the boundary
where one valley scatters into the other (indicated in green for positive
and red for negative).}
\end{figure}
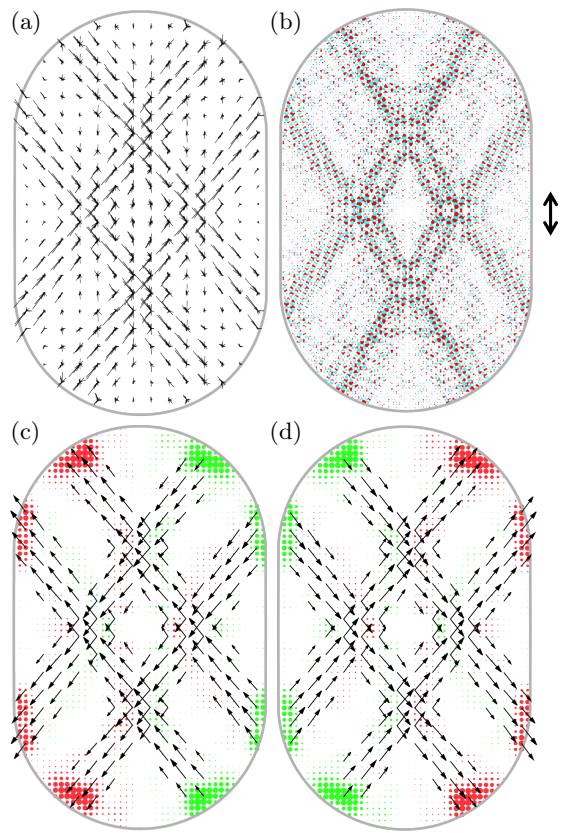

For the square lattice, time-reversal symmetry is expressed in the
Husimi projection by the fact that each Husimi vector is accompanied
by another of equal magnitude but opposite direction. This causes
the flux operator and Eq.~\ref{eq:Husimi-Vector-Old} to return null
results. The same is true for the honeycomb lattice, except that the
range of wavevectors available at low energies point towards the $K$
and $K^{\prime}$ valleys.

But when the Husimi vectors are weighted by the group-velocity and
not the wavevector, a different behavior emerges. In the honeycomb
lattice, group-velocity doesn't correlate at low energies with $\vec k$
but $\vec k-\vec K^{(\prime)}$. If one examines the Husimi projection
for each valley individually, it is no longer true that each Husimi
vector is accompanied by its opposite. Rather, each valley is the
time-reversal symmetric version of the other, allowing Husimi vectors
in each valley to sum to non-trivial results.

Fig.~\ref{fig:Graphene-Eig}a shows the Husimi map of the $\valley$
valley for a high-energy eigenstate in part (b) where the strong pull
towards the three preferred group-velocities is evident. In parts
(c) and (d), the multi-modal analysis for the $\valley$ and $\othervalley$
valleys are shown. According to the time-reversal symmetric relation,
the Husimi map for the $\othervalley$ valley is the precise inverse
of the $\valley$ valley. While the classical trajectories are evident
in the wavefunction (Fig.~\ref{fig:Graphene-Eig}b), the Husimi map
identifies their orientation for each valley.

Because Husimi vectors for each valley no longer sum to zero, it is
possible to produce divergence in the Husimi map for each wavevector.
This is identical to angular deflection in Eq.~\ref{eq:Summed-Divergence}
except that the absolute value is not taken in the integrand. And
like angular deflection, the integrand must be weighted by the group-velocity
vector, or non-trivial results emerge in the bulk of the system. Summing
the divergence for all wavevectors produces the total divergence,
which appears in green and red in Figs.~\ref{fig:Graphene-Eig}c
and \ref{fig:Graphene-Eig}d to indicate positive and negative values.
These points are, in fact, sources and drains for each valley, and
represent the inter-valley scattering points along the boundary, whose
scattering properties depend on the angle of the cut\cite{BeenakkerBC}.
The results in Fig.~\ref{fig:Graphene-Eig} suggest that each classical
trajectory in this wavefunction shares half of its existence in one
valley, and half in the other.

\subsection{Group Velocity Warping\label{sub:Group-Velocity-and}}

\begin{figure}
\begin{centering}
\begin{overpic}[width=0.85\columnwidth]{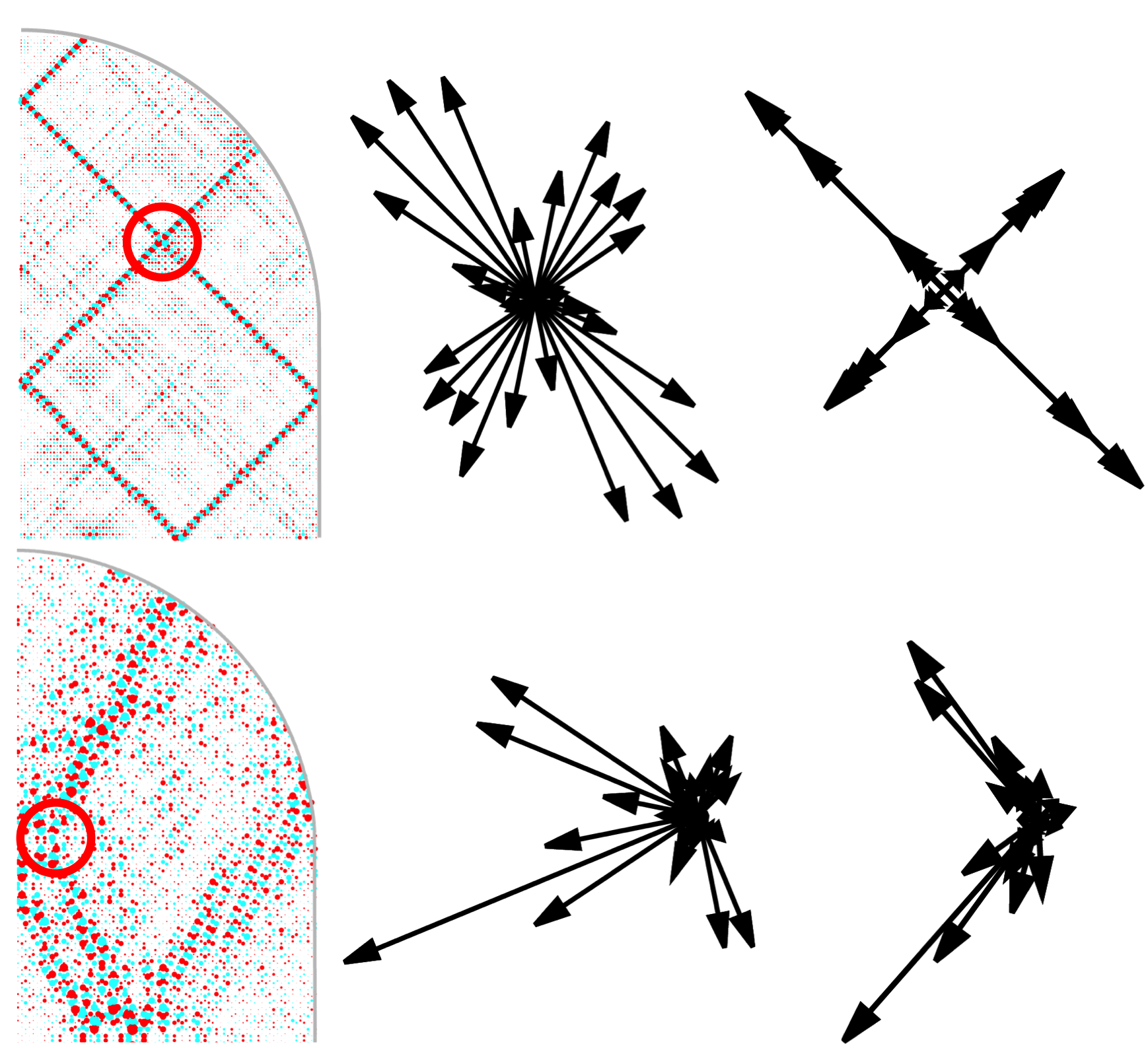}\put(-5,87){\figlab{a}}\put(-5,41){\figlab{b}}\end{overpic}
\par\end{centering}

\caption{\label{fig:Phase and Group Velocity}Two stadium eigenstates for the
square lattice (a) and the honeycomb lattice (b). The wavefunctions
(left), wavevector Husimi (middle) and group-velocity Husimi (right)
projections are shown for the points circled in red. Uncertainties
for both projections are $\Delta k/k=20\%$. Not only is there more
spread to the wavevector projections, these projections also indicate
markedly different trajectory paths than the group-velocity equivalents.
Moreover, the group-velocity projections are more consistent with
the paths indicated by the wavefunctions.}
\end{figure}

This section expands upon our findings in Figs.~\ref{fig:Billiard Eigenstates}
and \ref{fig:Graphene-Eig} by examining Husimi projections in detail.
In Fig.~\ref{fig:Phase and Group Velocity}, we show Husimi projections
for the square (a) and honeycomb (b) lattices in both wavevector and
group-velocity representations. As expected, the spread of each Husimi
projection is dramatically reduced in the group-velocity representation,
a consequence of group-velocity warping and consistent with Fig.~\ref{fig:Sunbursts}.
Moreover, a close examination reveals that Husimi wavevectors can
point along surprisingly divergent angles from their trajectories,
emphasizing the extent to which group-velocity warping establishes
such states.

If it is possible to produce similar classical trajectories using
a wider variety of wavevectors for lattices, then how are wavevectors
distributed in this wider range? We can provide an answer by summing
the Husimi projections over a range of eigenstates. We find that with
a sufficient range of eigenstates, neither wavevector nor group-velocity
distributions vary across the bulk of the system, except along the
boundaries. For the square-lattice billiards, directions parallel
to boundaries are emphasized, which is consistent with Dirichlet boundary
conditions. This occurs even for jagged edges that do not fall along
a symmetry axis of the underlying lattice.

Boundaries on the honeycomb lattice emphasize either parallel trajectories
for intra-valley scatterers (zig-zag) or perpendicular trajectories
for \emph{inter}-valley scatters (armchair). We find that honeycomb
lattice systems are more sensitive than the square lattice to the
set of states we sum over, requiring a larger sum to provide uniform
results. The distance from the edge where the emphasis occurs is a
function of the characteristic wavelength; for the square lattice,
this is $k$ and in the honeycomb lattice $q=\left|\vec k-\vec K'\right|$.
As a result, for small enough honeycomb systems, the emphasis of directions
along the boundary can persist into the bulk.

\begin{figure}
\begin{centering}
\begin{overpic}[width=0.85\columnwidth]{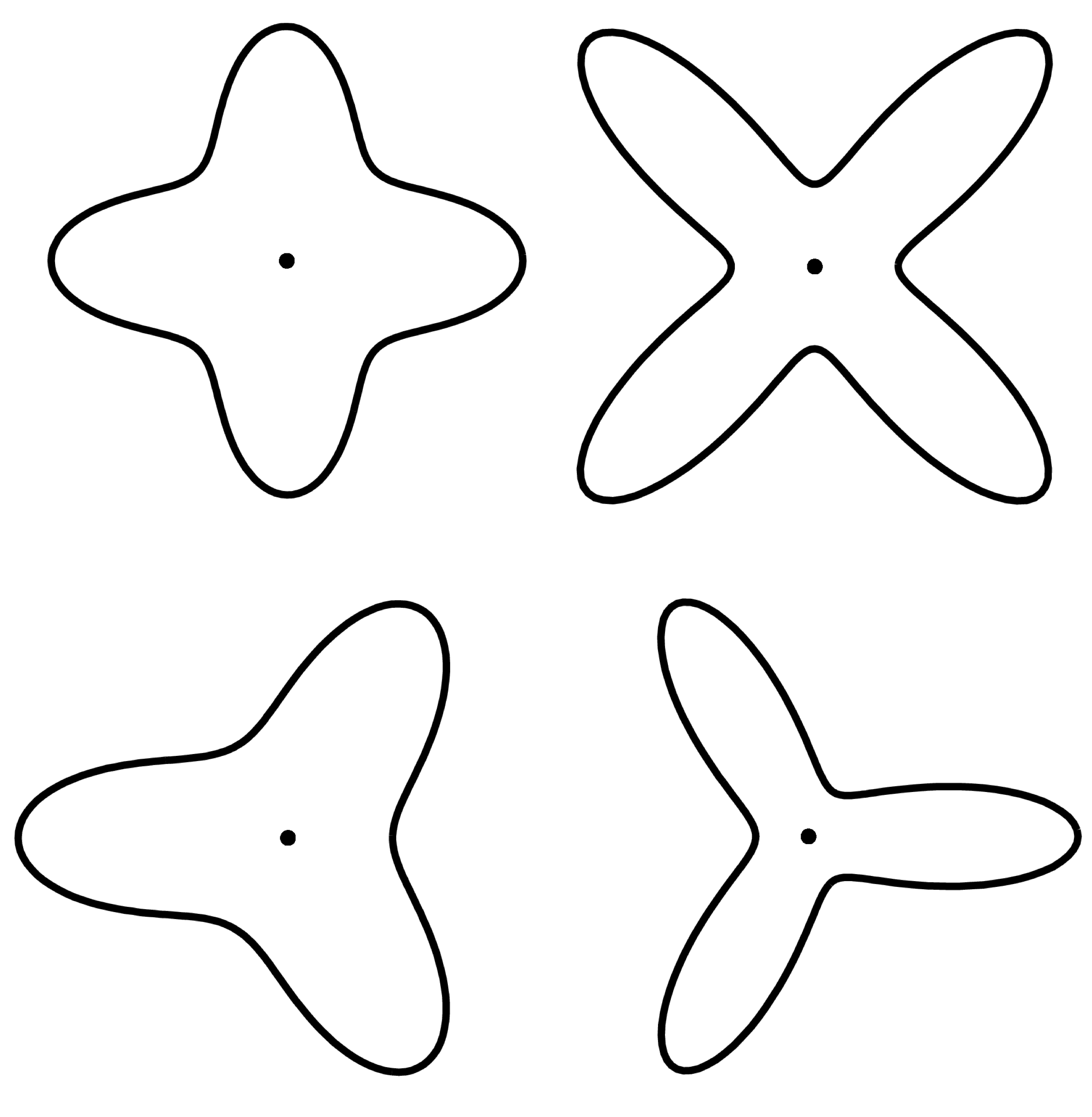}\put(5,95){\figlab{a}}\put(45,95){\figlab{b}}\put(5,42){\figlab{c}}\put(45,42){\figlab{d}}\end{overpic}
\par\end{centering}

\caption{\label{fig:Ergo}The distribution of Husimi vectors from the red circles
in Fig.~\ref{fig:Phase and Group Velocity}, summed over hundreds
of eigenstates near $E=3.5t$ for the square lattice and $E=0.8t$
for the honeycomb lattice, with a coherent wavepacket spread of $\Delta k/k=10\%$.
Above, the wavevectors in the square lattice (a), group-velocities
in the square lattice (b), wavevectors for the $\valley$ valley in
the honeycomb lattice (c), and group-velocities for the $\valley$
valley in the honeycomb lattice (d). Husimi projections tend to emphasize
wavevectors away from the preferred group-velocity directions (a,c),
but not enough to overcome that preference in the group-velocity distribution.}
\end{figure}
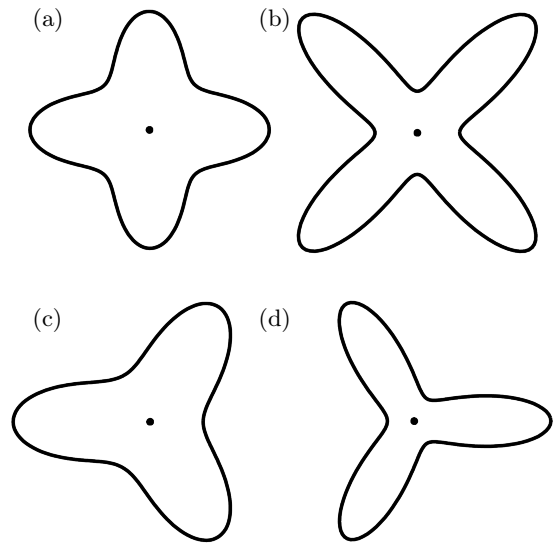
In Fig.~\ref{fig:Ergo}, we show a representative distribution of
the wavevectors and group-velocities at the points circled in red
in the stadium systems presented in Fig.~\ref{fig:Phase and Group Velocity},
using Husimi projections with a coherent spread of $\Delta k/k=10\%$.
The states used for the square-lattice system are near energies of
$E=3.5t$, and at $E=0.8t$ for the honeycomb lattice. More details
can be found in Appendix \ref{sec:ergo-appendix}. Fig.~\ref{fig:Ergo}
shows that the distribution among wavevectors emphasizes directions
\emph{away} from the preferred directions in group velocity. For certain
energy regimes, \emph{neither} wavevectors \emph{nor }group velocities
are evenly distributed across all eigenstates of a lattice system.

\subsection{Internal Bragg Diffraction\label{sub:Internal-Bragg-Diffraction}}

\begin{figure}
\begin{centering}
\includegraphics[width=0.85\columnwidth]{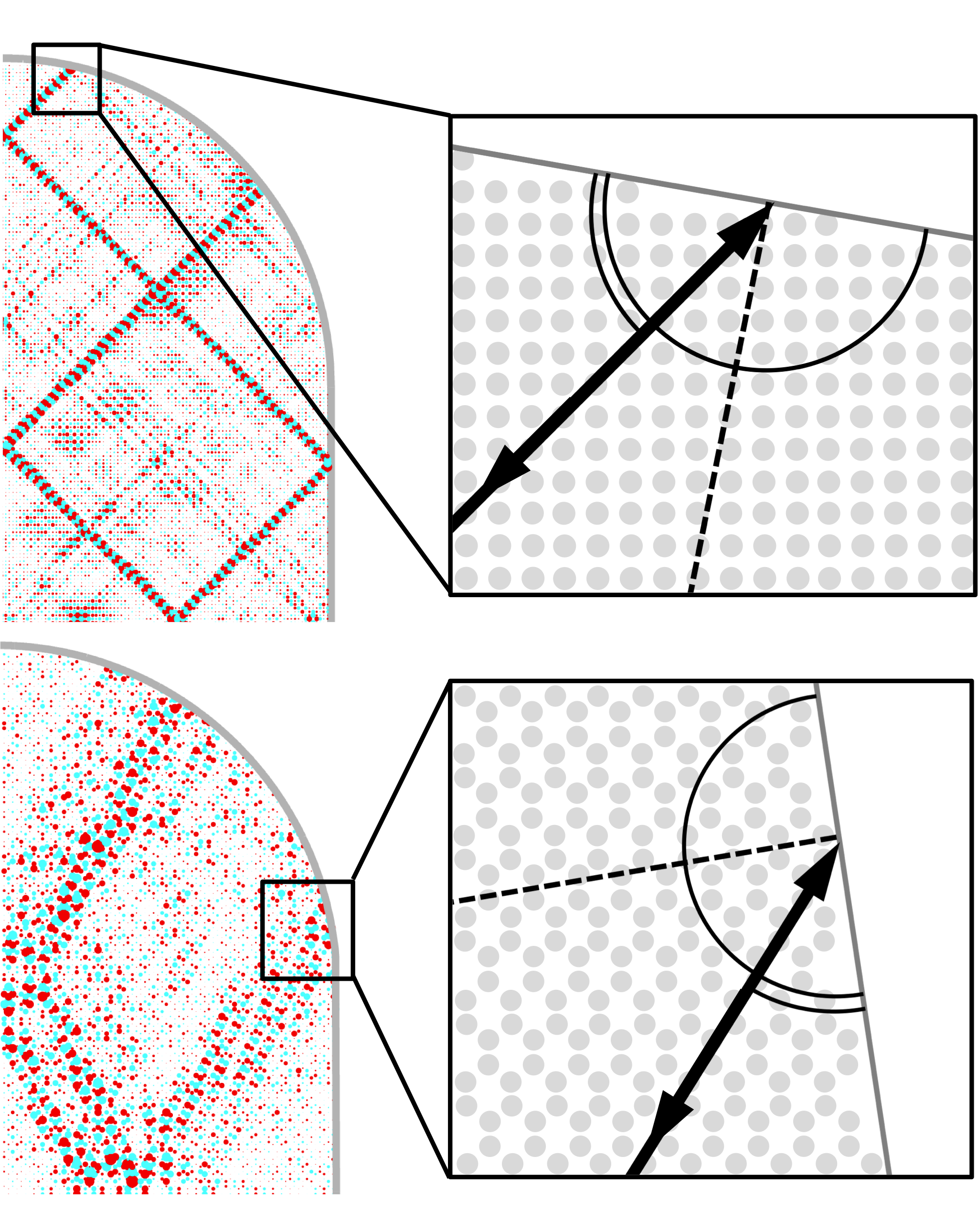}
\par\end{centering}

\caption{\label{fig:Schematics}High-energy states in the square (a) and honeycomb
(b) lattices can exhibit unusual behaviors, such as group-velocity
warping and non-specular boundary reflections. The former can be seen
in the wavefunction (left) by the restriction of trajectories to $45^{\circ}$
diagonals for the square lattice (a) and the $60^{\circ}$ diagonals
for the honeycomb lattice (b). Non-specular reflections are magnified
in the schematic (right). Even though the absolute incoming and outgoing
angles for each point are the same angle, their angles of incidence
(single and double arcs) are strongly divergent. }
\end{figure}
The high-energy eigenstates from Fig.~\ref{fig:Phase and Group Velocity}
exhibit an unusual behavior: the self-looping classical trajectories
that are strongly emphasized in the wavefunctions do not exhibit specular
reflection at the boundary. We clarify these reflections in the schematics
in Fig.~\ref{fig:Schematics}. Even though the absolute angles at
each reflection point fall along the same diagonal, the \emph{angles
of incidence} vary substantially between the incoming and outgoing
rays. In the honeycomb eigenstate (Fig.~\ref{fig:Schematics}b),
the reflection consists of scattering into the other valley and propagating
in the exact opposite direction.

While the reflections of many trajectories in high-energy states violate
specularity as a result of group-velocity warping, we have chosen
the states in Figs.~\ref{fig:Phase and Group Velocity} and \ref{fig:Schematics}
specifically because these reflections behave in unexpected ways.
Moreover, these surprising reflections occur only at certain points
along the boundary where the lattice cut deviates from an axis of
symmetry; specifically, they occur slightly off of clean cuts where
jaggedness is most prominent. 

\begin{figure}
\begin{centering}
\includegraphics[width=0.95\columnwidth]{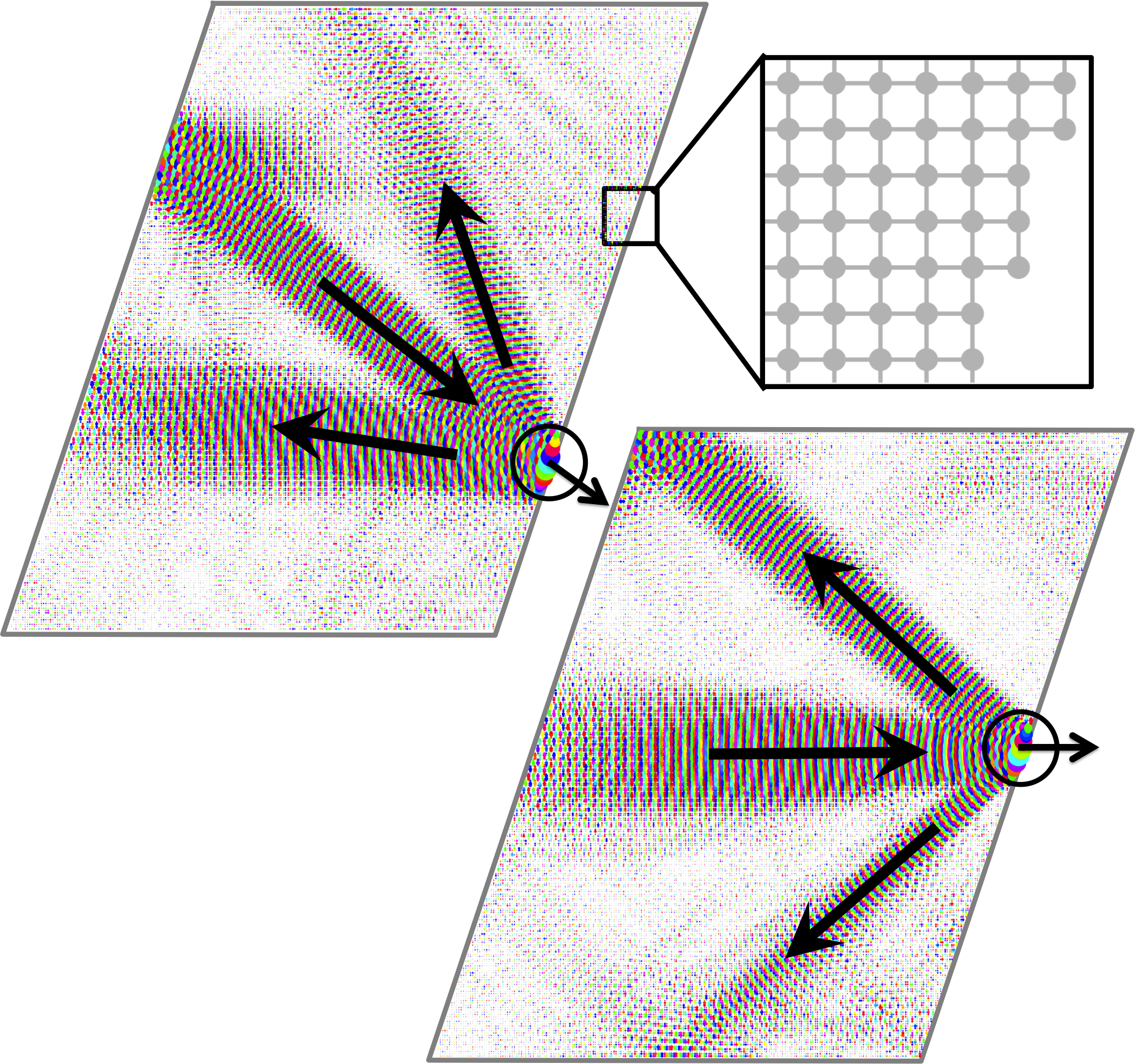}
\par\end{centering}

\caption{\label{fig:Gaussian Beam}Two Gaussian Beams, constructed by summing
the set of closed-system eigenstates in the energy range $2.48<E<2.52$
weighted by Eq.~\ref{eq:gaussian-beam}, using a coherent state with
momentum uncertainty $\Delta k/k=5\%$ that sits on the right-hand
boundary (black circles) with specified momentum (small black arrows).
The system is a square lattice cut at an $18^{\circ}$ angle (inset).
The incoming group-velocity angle is set to $0^{\circ}$ at top and
$-40^{\circ}$ at bottom.}
\end{figure}
To understand these unusual reflections, we use a technique called
the Gaussian beam\cite{quantum-optics}, which shows the entire set
of wavefunctions available to the system which intersect at a particular
point in both spatial and momentum coordinates. This is accomplished
by weighting the eigenstates $\left\{ \psi_{E}\right\} $ for a closed
system by a coherent state $\ket{\vec r_{0},\vec k_{0},\sigma}$ which
satisfies the dispersion relation at energy $E_{0}$. To examine reflections
at jagged boundaries, we place $\vec r_{0}$ along one of these boundaries
and $\vec k_{0}$ pointing away from the bulk. Each eigenstate is
associated with an eigenenergy $E$ so that the Gaussian beam $\Psi$
is defined as 
\begin{equation}
\Psi=\sum_{E}\braket{\psi_{E}}{\vec r_{0},\vec k_{0},\sigma}\psi_{E}.\label{eq:gaussian-beam}
\end{equation}
Because of the finite uncertainty of the coherent state, only wavefunctions
at energies close to $E_{0}$ contribute to the final result. Thus,
only a finite range centered around $E_{0}$ must be considered.

It is important to choose the spread of the coherent state wisely.
Too large a coherent state restricts the set of eigenstates that contribute
to the sum, giving unclear results. Too small a coherent state does
not provide enough information to resolve features of the beam. In
Fig.~\ref{fig:Gaussian Beam} a compromise is chosen at $\Delta k/k=5\%$,
which provides a sufficient range of eigenstates to construct a clear
beam.

The classical paths suggested by the Gaussian beams in Fig.~\ref{fig:Gaussian Beam}
must all travel through the position $\vec r_{0}$ with momentum $\hbar\vec k_{0}$
(Eq.~\ref{eq:gaussian-beam}) defined by the coherent state for each
beam. In both top and bottom figures, the coherent state lies along
the right-hand boundary, although the wavevectors for each coherent
state differs. Because the breadth of a coherent state grows in time
when it propagates, each beam focuses at the coherent state, and spread
from its center. In both Figs.~\ref{fig:Gaussian Beam}a and \ref{fig:Gaussian Beam}b,
a specular reflected beam is present, but an additional reflected
beam emerges as a result internal Bragg diffraction\cite{bragg}.
These additional reflections also appear at scattering points for
the square lattice in Fig.~\ref{fig:Schematics}a.

\begin{figure}
\begin{centering}
\includegraphics[width=0.75\columnwidth]{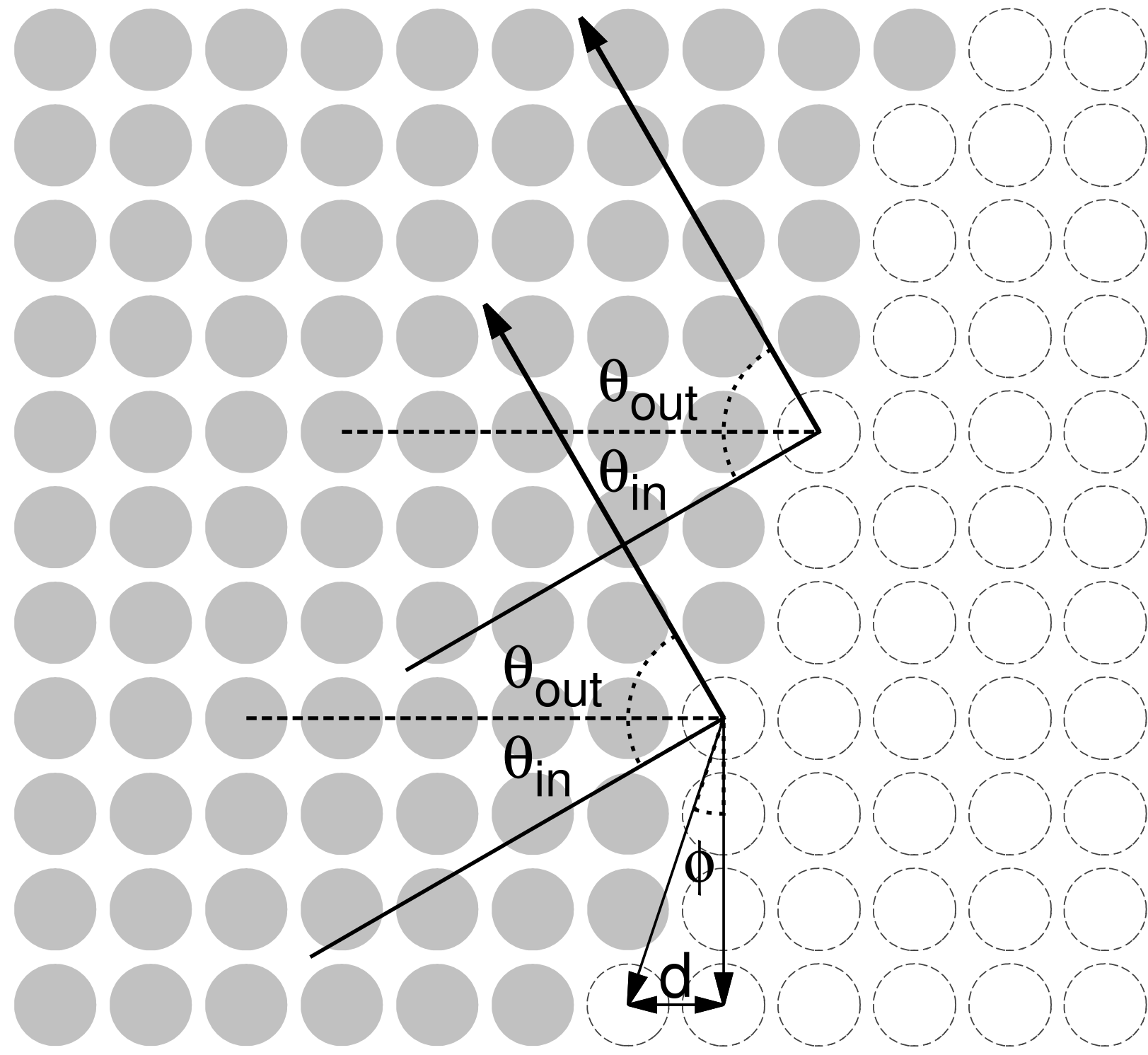}
\par\end{centering}

\caption{\label{fig:Reflection-Diagram}Schematic of internal Bragg diffraction.
In light gray are the lattice sites for the system, and in dashed
outlines are shown phantom lattice points outside the system where
the wavefunction must go to zero. A ray coming in at an angle $\theta_{\text{in}}$
and reflects at $\theta_{\text{out}}$ interferes with rays from adjacent
equivalent points on the boundary unit cell. The angle $\phi$ and
horizontal distance between adjacent unit cells $d$ are identical
to the boundary in \ref{fig:Gaussian Beam}.}
\end{figure}

It is possible to quantify internal Bragg diffraction by considering
an edge identical to the system in Fig.~\ref{fig:Gaussian Beam}
and depicted schematically in Fig.~\ref{fig:Reflection-Diagram}.
Here the boundary is cut at an angle $\phi\approx18^{\circ}$, where
$\phi=0$ is a vertical edge, and an incoming plane wave strikes the
surface at angle $\theta_{\text{in}}$, where $\theta_{\text{in}}=0$
points to the right, and positive angles point upward. This plane
wave reflects to an outgoing angle $\theta_{\text{out}}$ where $\theta_{\text{out}}=0$
points to the left and positive angles point upward.

If there is a repeating unit cell in the edge, two rays which hit
equivalent points of adjacent boundary unit cells gain or lose relative
phase based on their wavevectors and the different distances they
travel. For instance, a ray incurs an additional phase of $\delta=kd\frac{\sin\left(\theta-\phi\right)}{\sin\phi}$
when $\theta_{\text{in}}>-\phi$ and $\delta=kd\frac{\sin\left(\theta+\phi\right)}{\sin\phi}$
when $\theta_{in}\geq\phi$. Here $d$ is the horizontal distance
between identical points in adjacent unit cells and $k$ is the wavevector
magnitude of the incoming wave. When the plane wave is reflected,
its neighbor gains phase according to the above formulas, but with
$k$ indicating the outgoing wavevector magnitude. When these two
phases cancel or add to a multiple of $2\pi$, the two rays constructively
interfere. Since this is repeated over many unit cells, the interference
can be quite strong.

Because the wavelength shrinks with increasing energy, more Bragg
branches appear as energy goes up. And because the distance between
adjacent unit cells increases for slighter angles against an axis
of symmetry, more Bragg branches appear for shallower cuts. We find
that both of these criteria are satisfied for the reflection points
in Figs.~\ref{fig:Schematics} and \ref{fig:Graphene-Eig}.

\begin{figure}
\begin{centering}
\includegraphics[width=0.85\columnwidth]{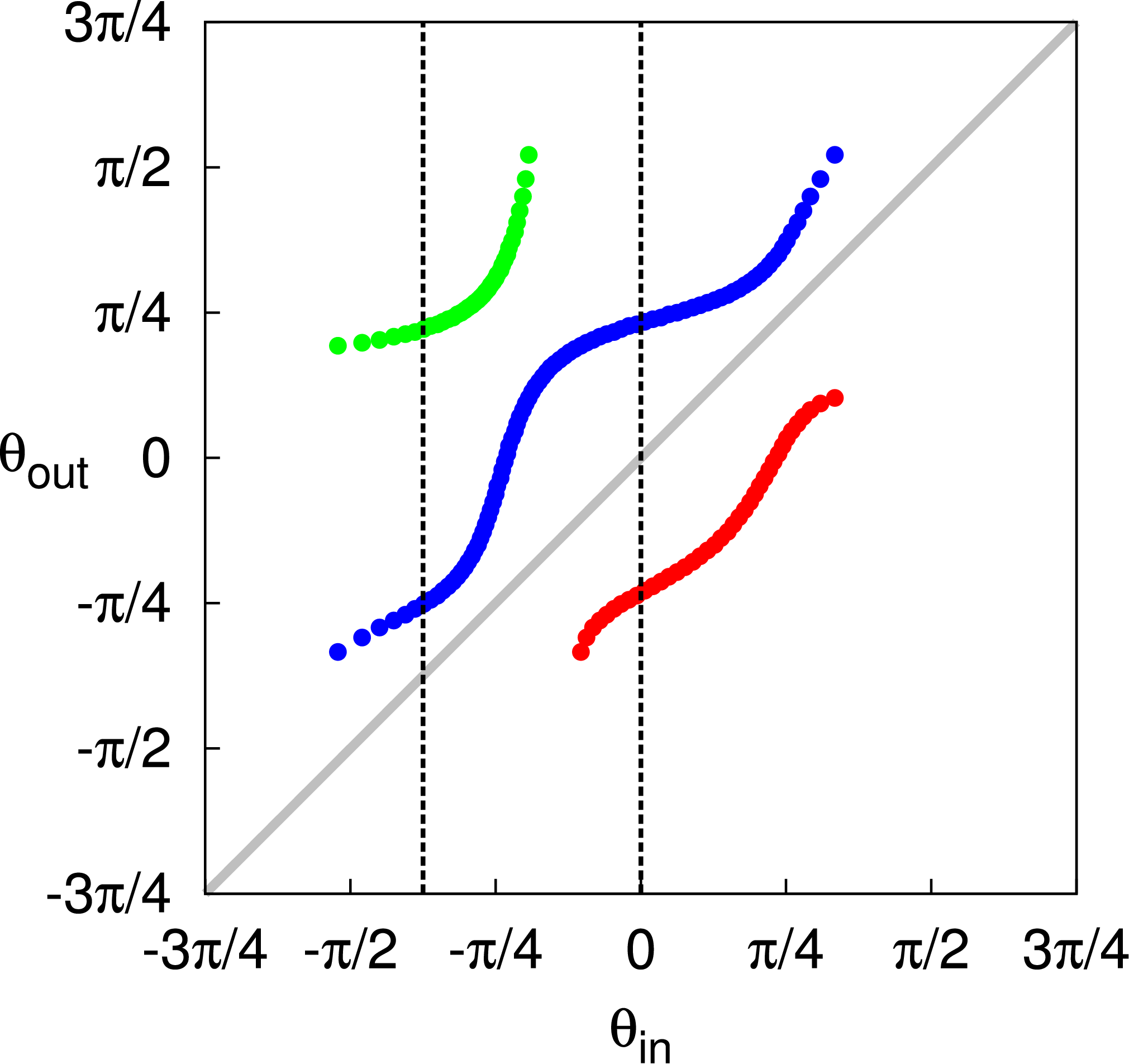}
\par\end{centering}

\caption{\label{fig:Gaussian Beam-Analysis}The internal Bragg relationship
for a square lattice with an $18{}^{\circ}$ cut as depicted in Figs.~\ref{fig:Gaussian Beam}
and \ref{fig:Reflection-Diagram}, computed using a scattering matrix
on a square-lattice with 50 vertical unit cells at energy $E=2.5t$.
The identity line is shown in grey. The two incoming group-velocity
angles from Fig.~\ref{fig:Gaussian Beam} of $0^{\circ}$ and $-40^{\circ}$
are shown in vertical black dashed lines. The specular line is shown
in blue, and the upper and lower branches are shown in green and red
respectively.}
\end{figure}

Using the above formulas, we can predict internal Bragg diffraction
for arbitrary cuts and energies. In Fig.~\ref{fig:Gaussian Beam-Analysis},
we present these results for the system in Fig.~\ref{fig:Gaussian Beam}.
The boundary unit cell consists of three vertical units and one horizontal
unit, so that $\phi\approx18^{\circ}$. The two incoming beams from
Fig.~\ref{fig:Gaussian Beam} are represented by vertical dashed
lines in Fig.~\ref{fig:Gaussian Beam-Analysis}. Each intersects
the graph at the three locations: along the identity line for the
incoming beam, along the blue specular line for an outgoing beam,
and along one of the Bragg branches for the other outgoing beam. Our
predictions are strongly validated by Fig.~\ref{fig:Gaussian Beam}.

The $\sigma$ spread of the test wavepacket used to create the Gaussian
beam only covers 4 steps along the cut, meaning that only a few surface
defects can produce substantial Bragg scattering. The ubiquity of
this effect has implications for ray-tracing methods, which bridge
classical and quantum explanations for phenomena such as fractal conductance
fluctuations\cite{Ray-Fractal-Conductance,Ray-Fractal2} and caustics\cite{Heller-Caustics,Ray-Berry-Caustics}
and encourages a re-examination ray-splitting\cite{ray-splitting}
and other hypothetical edge effects\cite{ray-edge-effects,ray-edge-diffraction}. 

Combining group-velocity restriction and internal Bragg diffraction,
we argue that the dense linear paths in the wavefunctions in Figs.~\ref{fig:Phase and Group Velocity}
and \ref{fig:Schematics} are indeed linked to classical rays which
bounce back and forth approximately linearly; at one boundary the
bounce is non-specular due to the cut of the edge and internal Bragg
diffraction. For the honeycomb lattice, each bounce can be additionally
associated with scattering into another valley. For both systems,
these wavefunction enhancements are not strictly scars\cite{PhysRevLett.53.1515},
which are generated by unstable classical periodic orbits in the analogous
classical limit (group velocity) system. Instead, the wavefunction
structures are likely normal quantum confinement to stable zones in
classical phase space.

\section{Conclusions}

We have expanded the vector Husimi projection technique, introduced
in Mason \emph{et al.}\cite{Mason-Husimi-Continuous}, from the continuous
system to lattices. We have demonstrated and explained two unusual
properties of lattice systems using the Husimi projection: group-velocity
warping and internal Bragg diffraction, both of which can strongly
affect the properties of classical dynamics of these systems, in particular
producing unexpected self-looping trajectories most visible in extreme-energy
states. We have also shown that Husimi projections can isolate multiple
bands which are simultaneously represented in the wavefunction, using
the two valleys of the honeycomb as an example. For the honeycomb
lattice, we have shown that one can identify locations of scattering
between valleys by measuring the divergence of the Husimi map for
each valley separately.
\begin{acknowledgments}
This research was conducted with funding from the Department of Energy
Computer Science Graduate Fellowship program under Contract No. DE-FG02-97ER25308.
MFB and EJH were supported by the Department of Energy, office of
basic science (grant DE-FG02-08ER46513).
\end{acknowledgments}
\appendix

\section{Wavevector and Group Velocity Distributions\label{sec:ergo-appendix}}

The plots in Fig.~\ref{fig:Ergo} are produced by summing the Husimi
vectors over many eigenstates of each system in Fig.~\ref{fig:Phase and Group Velocity}:
For the square lattice, 600 states with energies $3.46t<E<3.54t$
and for the honeycomb lattice, 300 states with energies $0.76t<E<0.84t$.
This is done for 256 wavevectors equally separated by angle, and then
a small Gaussian kernel is applied with angle width $\pi/32$. Each
Husimi vector is multiplied by the infinitesimal $dk$ determined
by the average distance to neighboring vectors in the sample, and
each calculation takes place at the points circled in red in Fig.~\ref{fig:Phase and Group Velocity}
with coherent spread of $\Delta k/k=10\%$. The contour line in the
dispersion relation is re-computed for each eigenstate to generate
the coherent states for the Husimi projection. This is done to ensure
that the steeper gradient of the dispersion relation near the preferred
group velocities does not affect our results.

\bibliographystyle{unsrt}

\end{document}